\documentclass[10pt]{article}
\usepackage{opex2}
\usepackage{amsmath,bm}
\usepackage[dvips]{graphicx}
\begin{document}

\title{Compensation for Extreme Outages caused by Polarization
Mode Dispersion and Amplifier noise}

\author{Vladimir Chernyak$^a$, Michael Chertkov$^b$, Igor
Kolokolov$^{b,c,d}$, and Vladimir Lebedev$^{b,c}$}

\address{$^a$Corning Inc., SP-DV-02-8, Corning, NY 14831, USA; \\
$^b$Theoretical Division, LANL, Los Alamos, NM 87545, USA; \\
$^c$Landau Institute for Theoretical Physics, Moscow, Kosygina 2, 117334, Russia; \\
$^d$Budker Institute of Nuclear Physics, Novosibirsk 630090, Russia.}

%\email{chertkov@lanl.gov}
%\homepage{http://www.opticsexpress.org/}

\begin{abstract}
Joint effect of weak birefringent disorder and amplifier noise on transmission in optical fiber
communication systems appears to be strong. The  probability of an extreme outage that corresponds to
anomalously large values of Bit Error Rate (BER) is perceptible. We analyze the dependence of the
Probability Distribution Function (PDF) of BER on the first-order and also higher-order PMD
compensation schemes.
\end{abstract}

% The commands, \begin{OEReferences} and \end{OEReferences}
% format the References section according to OpEx standard
% style, showing the title "References".
%
% The commands, \begin{OERefLinks} and \end{OERefLinks}
% format the References section according to OpEx standard
% style, if the references also include URLs or other
% unreviewed links.  In this case the title of the section
% is "References and unreviewed links".
%
\begin{OEReferences}
%\begin{thebibliography}{}
\bibitem{97PN} C. D. Poole and J. A. Nagel, in Optical Fiber
  Telecommunications, eds. I. P. Kaminow and T. L. Koch, Academic San
  Diego, Vol. IIIA, pp. 114, (1997).

\bibitem{99JNPG} R. M. Jopson, L. E. Nelson, G. J. Pendlock, and A. H. Gnauck,
 ``Polarization mode dispersion impairment in return to zero and non-return-tozero systems", in
 Tech. Digest Optical Fiber Communication Conf. (OFC'99), San Diego, CA, 1999, Paper WE3.

\bibitem{98Hei} F. Heismann, ECOC'98 Digest 2, 51 (1998).

\bibitem{00GK} J. P. Gordon and H. Kogelnik,
``PMD fundamentals: Polarization mode dispersion in optical fibers", PNAS {\bf 97}, 4541 (2000).

% 97PN,98Hei and 00GK are hree major (and modern) PMD reviews.

 \bibitem{94OYSE} T. Ono, S. Yamazaki, H. Shimizu, and H. Emura,
 ``Polarization control method for supressing polarization mode
 dispersion in optical transmission systems",
 J. Ligtware Technol. {\bf 12}, 891 (1994).

\bibitem{98HFW} F. Heismann, D. Fishman, and D. Wilson,
``Automatic compensation of first-order polarization mode dispersion in a 10
  Gb/s transmission system",
 in Proc. ECOC'98, Madrid, Spain, 1998, pp. 529-530.
% First order compensation possessing three degrees of freedom was
%experimentally demonstrated here.

\bibitem{99MK} L. Moller and H. Kogelnik,
``PMD emulator restricted to
  first and second order PMD generation",
 in PROC. ECOC'99, 1999, pp. 64-65
% It is shown here that first- and second- order PMD compensation
% can be achieved with a filter design requiring at least five degrees of freedom.

\bibitem{00BBBBW} H. B\"ulow, F. Buchali, W. Baumert, R. Ballentin, and T. Wehren,
  ``PMD mitigation at 10 Gbit/s using linear and nonlinear
  integrated electronic circuits",
  Electron. Lett. {\bf 36}, 163 (2000).
  An electrical PMD filter with nine adjustable parameters was tested.

%\bibitem{99NSYHMSGGSFWH} R. Noe, D. Sandel, M. Yoshida-Dierolf, S.
%  Hinz, V. Mirvoda, A. Schopflin, C. Glingener, E. Gottwald, C.
%  Scheerer, G. Fischer, T. Weyrauch, W. Haase,
  %``Polarization mode
  %dispersion compensation at 10,20, and 40 Gb/s with various optical equalizers",
%  J. Lightware Technol. {\bf 17}, 16002 (1999).
% An optical higher order PMD compensator with 73 control parameters was described here.

\bibitem{86PW}
C. D. Poole and R. E. Wagner, ``Phenomenological approach to polarization dispersion in long
single-mode fibres", Electronics Letters {\bf 22}, 1029 (1986).
% The term ``principal state'' was introduced here.

\bibitem{88Pol}
C. D. Poole, ``Statistical treatment of polarization dispersion in single-mode fiber'', Opt. Lett.
{\bf 13}, 687 (1988); {\bf 14}, 523 (1989).

\bibitem{91PWN}
C. D. Poole, J. H. Winters, and J. A. Nagel, ``Dynamical equation for polarization dispersion",
Opt. Lett. {\bf 16}, 372 (1991).

\bibitem{98Bul}
H. B\"{u}low, ``System outage probability due to first- and second-order PMD", IEEE Phot. Tech.
Lett. {\bf 10}, 696 (1998).
% Effect of second-order in H for pulse broadening was addressed here -
% generalization of Poole results for joint PDF of first and second-order dispersion
% vectors.

\bibitem{00KNGJ}
H. Kogelnik, L. E. Nelson, J. P. Gordon, and R. M. Jopson, ``Jones matrix for second-order
polarization mode dispersion", Opt. Lett. {\bf 25}, 19 (2000).

\bibitem{00ELMYT}
A. Eyal, Y. Li, W. K. Marshall, A. Yariv, and M. Tur, ``Statistical determination of the length
dependence of high-order polarization mode dispersion", Opt. Lett. {\bf 25}, 875 (2000).
%The two papers contain some extension of what Bulow has introduced.

\bibitem{94Des}
E. Desurvire, ``Erbium-Doped Fiber Amplifiers'', John Wiley \& Sons, 1994.

%\bibitem{01XSKA}
%C. Xie, H. Sunnerud, M. Karlsson, and P. A. Andrekson,
%``Polarization-Mode Dispersion-Induced Outages in Soliton Transmission Systems",
%IEEE Phot. Tech. Lett. {\bf 13}, 1079 (2001).
% PDF of BER from numerics is plotted in lin-log
% if to replot it in log-log the power-tail
% is cleraly seen. Funny enough, this major observation is
% not discussed in the paper at all
% - the plot is just used to contrast the reasonably performed
% linear case against the badly working
% nonlinear case.

%\bibitem{02SXKS}
%H. Sunnerud, C. Xie, M. Karlsson, and R. Samuelsson, ``A comparison between different PMD
%compensation techniques", Journal of Lightwave Technology, {\bf 20}, 368 (2002).
% Here we have some plot of the Log of Outage probability for
%numerical simulation as a function of
%$D_m z$ (for fixed values of $D_\xi$ and $z$).
%For $D_m z\ll 1$ the dependence is reasonably close
%to linear - that is our theory prediction.

%\end{thebibliography}
\end{OEReferences}

\noindent

{\bf Introduction:} Polarization Mode Dispersion (PMD) is recognized to be a substantial impairment
for optical fiber systems with the $40$-Gbs/s and higher transmission rates. One may not have a
complete control of PMD since the fiber system birefringence is changing substantially under the
influence of environmental condition (e.g., stresses and temperature) fluctuations, see e.g.
\cite{97PN,99JNPG}. Thus, dynamical PMD compensation became a major issue in modern fiber optics
communication technology \cite{98Hei,00GK}. Development of experimental techniques capable of the
first- \cite{94OYSE,98HFW,99MK} and higher-orders \cite{99MK,00BBBBW} PMD compensation have raised
a question of how to evaluate the compensation success (or failure). Traditionally, the statistics
of the PMD vectors of first \cite{86PW,88Pol,91PWN} and higher orders \cite{98Bul,00KNGJ,00ELMYT}
is considered as a measure for any particular compensation method performance. However, these
objects are only indirectly related to what actually represents the fiber system reliability. In
this letter we show that the PMD effects should be considered jointly with impairments due to
amplifier noise, since fluctuations of BER caused by variations of the birefringent disorder, are
substantial. We demonstrate that probability of extreme outages is much larger than one could
expect from naive estimates singling out effects of either of the two impairments. This phenomenon
is a consequence of a complex interplay between the impairments of different natures. (Birefringent
disorder is frozen, i.e. it does not vary on all propagation related time scales, while the
amplifier noise is extremely short-correlated.) The effect may not be explained in terms of just
an average value of BER, or statistics of any PMD vectors of different orders, but rather should
be naturally described in terms of the PDF of BER, and specifically its tail. A consistent
theoretical approach to calculating the tail will be explained briefly, with a prime focus on the
analysis of the first- and higher-order compensation effects on the extreme outages measured in
terms of the PDF of BER.

{\bf Bit-Error-Rate:} We consider the so-called return-to-zero modulation format, when pulses
(information carriers) are well separated in time, $t$. The quantity measured at the output of the
optical fiber line is then pulse intensity:
 \begin{equation}
 I=\int \mathrm dt\, G(t)
 \left|{\cal K}\bm\varPsi(Z,t)\right|^2 \,,
 \label{nnn} \end{equation}
where $G(t)$ is the convolution of the electrical (current) filter function with the sampling
window function. The two-component complex field $\bm\varPsi(Z,t)$ describes the output signal
envelope. The two components correspond to two polarizations of the optical fiber mode. The linear
operator ${\cal K}$ in Eq. (\ref{nnn}) stands for a variety of engineering ``tricks" applied to the
output signal. They consist of the optical filter ${\cal K}_f$, and the compensation ${\cal K}_{c}$
parts, respectively,  assuming the compensation is applied first followed by filtering, i.e.
${\cal K}={\cal K}_f\times{\cal K}_c$. Ideally, $I$ accepts two different values depending on
whether the information slot is vacant or filled. However, the impairments enforce deviations of
$I$ from the fixed values. Therefore, one has to introduce a threshold (decision level) $I_d$ and
declare that the signal encodes ``1" if $I>I_d$ and is related to ``0" otherwise. Sometimes the
information is lost, i.e. an initial ``1" is detected as a ``0" at the output or vise versa. BER is
the probability of such ``error" event (with statistics being collected over many pulses coming
through a fiber with a given realization of birefringent disorder). For successful system
performance the BER must be extremely small, i.e. both impairments typically cause only small
distortions to a pulse. It is straightforward to verify that anomalously high values of BER
originate solely from the ``$1\to0$" events. We denote the probability of such events by $B$.
Non-zero $B$ are caused by the noise, the values, however, depending on particular realization of
the birefringent disorder.

{\bf Noise averaging:} We consider the linear propagation regime, when the output signal
$\bm\varPsi$ can be decomposed into two contributions: $\bm\varphi$, related to a noiseless
initial pulse evolution and the noise-induced $\bm\phi$  part of the signal. $\bm\phi$
appears to be a zero-mean Gaussian variable (insensitive to a particular birefringence and chromatic
dispersion in the fiber) and is completely characterized by the pair correlation function
 \begin{equation}
 \langle\phi_\alpha(Z,t_1)\phi^\ast_\beta(Z,t_2)\rangle
 =D_\xi Z\delta_{\alpha\beta} \delta(t_1-t_2).
 \label{phiphi} \end{equation}
Here, $Z$ is the total length of the fiber line, and the product $D_\xi Z$ is the amplified
spontaneous emission (ASE) spectral density accumulated along the line. The
coefficient $D_\xi$ is introduced into Eq. (\ref{phiphi}) to reveal the linear growth of the ASE
factor with $Z$ \cite{94Des}.

{\bf Disorder averaging:} The noise-independent part of the signal is
 \begin{eqnarray}
 \bm\varphi=e^{i\eta\partial_t^2}
 \hat U \bm\varPsi_0(t) \,, \quad
 \hat U= T\!\exp\left[\int_0^Z\!\!\!\mathrm dz\,
 \hat m(z)\partial_t\right],
 \label{tu} \end{eqnarray}
where $\bm\varPsi_0(t)$, $\eta=\int_0^Z\!\!\!\mathrm dz\,d(z)$, $z$, and $d(z)$
are the input signal profile, the integral chromatic dispersion, coordinate along the fiber,
and the local chromatic dispersion, respectively. The ordered exponent $\hat U$ depends on the
$2\times2$ matrix $\hat{m}(z)$ that characterizes the birefringent
disorder. The matrix can be represented as $\hat m=h_j\hat{\sigma}_j$,
$h_j(z)$ being a real three-component field and $\hat{\sigma}_j$
the Pauli matrices. Averaging over many states of the birefringent
disorder any given fiber is going through (birefringence varies on a
time scale much longer than any time scale related to
the pulse propagation through the fiber), or over instant states of
birefringence in different fibers, one finds that $h_j(z)$ is a zero-mean Gaussian
field described by the following pair correlation function
 \begin{equation}
 \langle h_i(z_1)h_j(z_2)\rangle
 =D_m\delta_{ij}\delta(z_1-z_2).
 \label{hh} \end{equation}
If birefringent disorder is weak the integral $\bm H=\int_0^Z\mathrm dz\,\bm h(z)$
coincides with the PMD vector. Thus, $D_m=k^2/12$, where $k$ is the so-called PMD coefficient.

{\bf PDF of BER} is the proper object to describe fluctuations of $B$ caused by the birefringent
disorder. For successful fiber system performance the BER should be extremely small, i.e.
typically both impairments can cause only small distortions of a pulse. Stated differently, the
optical signal-to-noise ratio (OSNR) and the ratio of the squared pulse width to the mean square
value of the PMD vector are both large. OSRN can be estimated as $I_0/(D_\xi Z)$ where
$I_0=\int\mathrm dt\,|\varPsi_0(t)|^2$ is the initial pulse intensity, and the integration goes
over a single slot populated by an ideal (initial) pulse, encoding ``1''. Since the value of OSNR
is large averaging over the noise can be performed using the saddle-point method. This leads to a
conclusion that $D_\xi Z \ln B$ depends on the birefringence, shape of the initial signal and the
details of the compensation and measurement procedures, being, however, independent of the noise.
Typically, $B$ fluctuates around $B_0$, the zero-disorder, $h_j=0$, value of $B$. For any finite
value of $\bm h$ one gets, $\ln(B/B_0)=\Gamma I_0/(D_\xi Z)$, where the dimensionless factor
$\Gamma$ depends on $\bm h$. Since the noise is weak, even  small disorder can generate strong
increase in $B$. This implies that a perturbative calculation of $\Gamma$ based on expanding the
ordered exponent $\hat U$ in Eq. (\ref{tu}) in powers of $\hat m$, describes the most essential
part of the PDF of $B$.  Thus, in the situation when no compensation is applied one derives
$\Gamma=\mu_1 H_3/b$, whereas in the simplest case of the ``setting the clock" compensation,
accounting for the average (typical) temporal shift, one arrives at
$\Gamma=\mu_2(H_1^2+H_2^2)/b^2$, $b$ being the pulse width and $\mu_{1,2}$ being dimensionless
coefficients.

{\bf Long tail:} The PDF of $B$, ${\cal S}(B)$ (that appears in the result of averaging over many
realizations of the birefringent disorder) can be found by recalculating the statistics of $H_j$
using Eq. (\ref{hh}) followed by substituting the result into the corresponding expression that
relates $B$ to $H_j$. Our prime interest is to describe the PDF tail that corresponds to the
values of $H_j$ substantially exceeding their typical value $\sqrt{D_m Z}$ remaining, however,
much smaller than the signal duration $b$. In this range one gets the following estimates for the
differential probability ${\cal S}(B)\, \mathrm d B$:
 \begin{equation}
 a)\ \exp\left[-\frac{D_\xi^2Z b^2}{2D_m\mu_1^2I_0^2}
 \ln^2\left(\frac{B}{B_0}\right)\right]
 \frac{dB}{B}, \quad b)\
 \frac{B_0^\alpha\,\mathrm d B}{B^{1+\alpha}}\,,
 \label{PDFBER} \end{equation}
where (a) corresponds to the no-compensation situation, (b) stands for the optimal
``setting the clock" case,
and $\alpha\equiv D_\xi b^2/(2\mu_2D_m I_0)$. Note, that the result in the case (b)
shows a steeper decay compared to the case (a), which is a natural consequence of the ``setting the
clock" compensation.

{\bf PMD compensation:} One deduces from Eqs. (\ref{nnn},\ref{tu}) that the output intensity
depends on the birefringent disorder via the factor ${\cal K}_c\hat U$. The idea of the
compensation can be restated in a more formal way as building a linear operator ${\cal K}_c$ that
suppresses the dependence of ${\cal K}_c\hat U$ on $\bm h$. The so-called first-order compensation
${\cal K}_c={\cal K}_1$
 \begin{equation}
 {\cal K}_1=\exp\left(-\int_0^Z\!\!\! \mathrm dz\, h_j
 \hat\sigma_j\partial_t\right) \,,
 \label{kk1} \end{equation}
boils down to compensating the first term in the expansion of the ordered exponential $\hat{U}$  in
$\bm h$ \cite{98Bul,00KNGJ,00ELMYT}. Technically, this is achieved by sending the signal aligned
with either of the two principal polarization states of the fiber \cite{94OYSE}, or inserting a PMD
controller (a piece of polarization maintaining fiber with uniformly distributed well-controlled
birefringence) at the receiver \cite{98HFW}. Expanding ${\cal K}_1 \hat U$ in $\bm h$ followed by
substituting the result into Eq. (\ref{nnn}) and evaluating $B$ leads to
 \begin{eqnarray}
 \Gamma = (\mu'_2/b)^2\int_0^Z\!\!\!\mathrm
 d z'\int_0^{z'}\!\!\!\mathrm d z
 \left[h_1(z')h_2(z)\!-\!h_2(z')h_1(z)\right],
 \label{gmu3} \end{eqnarray}
$b$ being the pulse width, and only the leading $O(h^2)$ term is retained in Eq. (\ref{gmu3}).
The dimensionless coefficient $\mu'_2$ is related to the output signal chirp,
produced by the initial signal chirp and/or the nonzero integral
chromatic dispersion $\eta$. Recalculating the statistics of $\Gamma$ using Eqs. (\ref{hh},\ref{gmu3})
one obtains the following tail ${\cal S}(B)$ for the PDF of $B$
 \begin{equation}
 {\cal S}(B)\,\mathrm d B
 \sim\frac{B_0^\gamma\,\mathrm d B}
 {B^{1+\gamma}}, \qquad
 \gamma\equiv\frac{\pi D_\xi b^2}{2|\mu'_2| D_m I_0},
 \label{power3} \end{equation}
and Eq. (\ref{power3}) holds when $\ln(B/B_0)\gg|\mu'_2|D_m I_0/[D_\xi b^2]$.

{\bf Non-chirped signal:} If the output signal is not chirped $\mu'_2=0$ and the first
non-vanishing term in the expansion of $\Gamma$ in $h_j$ is of the third order. Expanding ${\cal
K}_1 \hat U$ up to the leading (third order) term yields
 \begin{eqnarray}
 \Gamma=\frac{\mu_3}{b^3}\!
 \int_0^Z\!\!\!\!\mathrm d z_1
 \!\int_0^{z_1}\!\!\!\!\mathrm d z_2
 \!\int_0^{z_2}\!\!\!\mathrm d z_3\,
 \Bigl\{2h_3(z_1){\cal H}(z_2,z_3)
 %\nonumber \\
 \!-\!h_3(z_2){\cal H}(z_1,z_3)
 \!-\!h_3(z_3){\cal H}(z_1,z_2) \Bigl\} ,
 \label{igor} \end{eqnarray}
with ${\cal H}(z_1,z_2)=h_1(z_1)h_1(z_2)+h_2(z_1)h_2(z_2)$. Substituting Eq. (\ref{igor}) into the
expression for $B$ in terms of $\Gamma$ and making use of Eq. (\ref{hh}) leads to a representation
of the PDF of $B$ as a path-integral over $\bm h$. Integrating over $h_3$ explicitly and
approximating the resulting integral over $h_{1,2}$ by its saddle-point value, one finds the PDF
tail
 \begin{equation}
 \ln{\cal S}\approx -4.2
 \frac{(D_\xi Z/I_0)^{2/3}b^2}{\mu_3^{2/3}D_m Z}
 \left(\ln\frac{B}{B_0}\right)^{2/3}.
 \label{igor2} \end{equation}
Eq. (\ref{igor2}) is valid at $D_\xi Z\ln(B/B_0)\gg \mu_3(D_mZ)^{3/2}I_0/b^3$.

\begin{figure}
 \centerline{\includegraphics[width=0.9\textwidth]{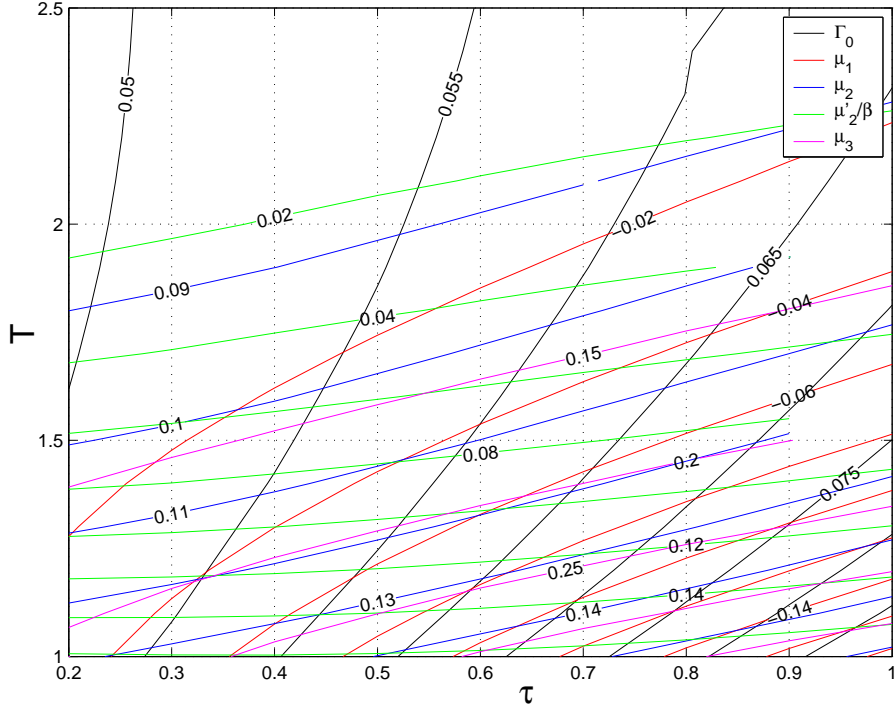}}
 \caption{Dependence of $\Gamma_0=- D_\xi z\ln B_0/I_0$, $\mu_1$, $\mu_2$, $\mu'_2/\beta$ and $\mu_3$ on $T,\tau$,
 for the model explained in the text. All quantities are
 measured in the units corresponding to the pulse width and amplitude both equal to one.}
 \label{mu123} \end{figure}

{\bf Simple model:} The dimensionless coefficients $\mu'_2$, $\mu_3$ can be computed in the
framework of a simple model, with the decision level threshold $I_d$ being twice smaller than the
ideal intensity, the Lorentzian profile of optical filter, ${\cal K}_f \bm\varPsi=\int_0^t\mathrm
dt'\exp[-t/\tau] \bm\Psi(t-t')/\tau$, and the step function form for $G$, $G(t)=\theta(T-|t|)$. We
also consider a Gaussian weakly-chirped initial signal
$\bm\varPsi_0\propto\exp(-t^2/[2b^2])(1+i\beta_{in} t^2/b^2)$, $\beta_{in}\ll 1$ (here both the
signal amplitude and its width are re-scaled to unity). The  output signal chirp becomes
$\beta=\beta_{in}+\eta$, $\eta$ being the integral dimensionless chromatic dispersion. Then,
$\mu'_2$ is proportional to $\beta$, and the slope $\mu'_2/\beta$, found from the saddle-point
equations numerically, along with corresponding values of $\Gamma_0\equiv- D_\xi z\ln B_0/I_0$ and
$\mu_{1,2,3}$ are shown in Fig. \ref{mu123} for a reasonable range of the parameters $T,\tau$
(measured in the units of the pulse width $b$).

{\bf Advanced compensation:} The fiber system performance can be improved even further.
First of all, special filtering efforts can enforce the output pulse symmetry under the
$t\to-t$ transformation. Then the $O(H^3)$ contribution to $\Gamma$
will also be cancelled out and Eq. (\ref{igor})
will be replaced by $\Gamma=O(H^4)$. Second, one can use a more sophisticated
compensation ${\cal K}_c$ aiming to cancel as much terms in the expansion of
${\cal K}_c \hat U$ in $\bm h$ as possible. This approach
corresponds to the so-called high-order compensation techniques
implemented experimentally in many modern setting (see e.g.,
\cite{99MK,00BBBBW}). The high-order compensation can
substantially reduce the dependence of $\Gamma$ on $\bm h$, leading to
$\Gamma\sim \mu_k (H/b)^{k}$ (where $k$ exceeds by one the compensation
degree if no additional cancellations occur). In this case
the tail of the PDF of $\Gamma$ is estimated by $\exp[-(b^2/D_m
Z)(\Gamma/\mu_k )^{2/k}]$. This results in the following expression
for the tail of the PDF of $B$,
 \begin{equation}
 \ln{\cal S}\sim -(b^2/D_m Z)
 [\mu_k^{-1}D_\xi Z\ln(B/B_0)/I_0]^{2/k},
 \label{korder} \end{equation}
valid for $D_\xi Z \ln(B/B_0)\gg (D_m Z/\mu_k b^2)^{k/2}I_0$. Eq. (\ref{korder}) generalizes
Eqs. (\ref{power3},\ref{igor2}). One concludes that, as anticipated, the compensation does
suppress the PDF tail. One also finds the outage probability ${\cal O}$, defined as
${\cal O}=\int^1_{B_\ast}\mathrm dB\,{\cal S}(B)$ (with $B_\ast$ being some fixed value much
larger than $B_0$), $\ln{\cal O}\sim-[(\mu_kI_0)^{-1}D_\xi Z\ln(B_\ast/B_0)]^{2/k}b^2/(D_mZ)$.

{\bf Example:} Summarizing, our major result is quantitative description of the suppression of the
extremely long tail in the PDF of BER. We find it useful to conclude with presenting a numerical
example that corresponds to a case relevant for the optical fiber communications. Consider a fiber
line with $\Gamma_0=0.06$, $\mu_1=0.06$, $\mu_2=0.12$, $\mu'_2=0.15$ and $\mu_3=0.35$, which is
also characterized by typical bit-error probability, $B_0=10^{-12}$ that corresponds to
$I_0/[D_\xi Z]\approx460$. We also assume that the PMD coefficient, $k=\sqrt{12D_m}$, is $0.2$
$ps/\sqrt{km}$, the pulse width is $b=25$ $ps$, and the system length is $Z=2,500$ $km$, i.e. $D_m
Z/b^2\approx 0.013$. Then the outage probability corresponding to $B_\ast=10^{-10}$, that is the
probability for $B$ to be at least $2$ orders of magnitude larger than $B_0$, is ${\cal
O}\approx0.35$ if no compensation is applied, see Eq. (\ref{PDFBER}a), while one derives ${\cal
O}\approx 0.04$, ${\cal O}\approx 4\cdot10^{-4}$ and ${\cal O}\approx 2\cdot10^{-13}$ for Eq.
(\ref{PDFBER}b), Eq. (\ref{power3}) and Eq. (\ref{igor2}), describing the cases of the ``setting
the clock", first- and second-order compensations, respectively.

{\bf Acknowledgment:} We are thankful to I. Gabitov for numerous valuable discussions. We also wish
to acknowledge the support of LDRD ER on ``Statistical Physics of Fiber Optics Communications" at
LANL.

\end{document}